\documentclass[doublecol]{epl2} 

\title{Particle shape dependence in {2D} granular media}
\shorttitle{Particle shape} 

\author{CEGEO\footnote{Collaborative group ``Changement d'Echelle dans 
les GEOmat\'eriaux'' (scale change in geomaterials)}, 
B. Saint-Cyr\inst{1,4} \and K. Szarf\inst{2} \and C. Voivret\inst{{5}} \and E. Az\'ema\inst{1} \and 
V. Richefeu\inst{2} \and J.-Y. Delenne\inst{{6}} \and G. Combe\inst{2} \and C. Nouguier-Lehon\inst{3} \and
P. Villard\inst{2} \and P. Sornay\inst{4} \and M. Chaze\inst{3} \and F. Radjai\inst{1} }
\shortauthor{CEGEO \etal}

\institute{                    
  \inst{1} University Montpellier 2, CNRS, LMGC UMR 5508, Place Eug\`ene Bataillon, F-34095 Montpellier Cedex, France.\\
  \inst{2} UJF-Grenoble 1, Grenoble-INP, CNRS UMR 5521, 3SR Lab., B.P. 53, F-38041 Grenoble Cedex 09, France.\\
  \inst{3} University of Lyon, Ecole Centrale de Lyon, LTDS UMR CNRS 5513, 36 avenue Guy de Colongue, F- 69134 Ecully cedex, France. \\
  \inst{4} CEA, DEN, SPUA, LCU, F-13108 Saint Paul Lez Durance, France.\\
  \inst{5} {SNCF Innovation and Research 
Immeuble Lumi\`ere, 40 Avenue des Terroirs de France, F-75611 Paris cedex 12, France.} \\
 \inst{6} {IATE, UMR 1208 INRA-CIRAD-Montpellier 
 Supagro-UM2, 2 place Pierre Viala, F-34060 Montpellier cedex 01, France.}
}

\abstract{
Particle shape is a key to the space-filling and strength 
properties of granular matter. We consider a 
shape parameter $\eta$   
describing the degree of distortion from a perfectly 
spherical shape. Encompassing most specific shape characteristics 
such as elongation, angularity and 
nonconvexity, $\eta$  is  a low-order but generic parameter that we used  
in a numerical benchmark test for a systematic investigation of  
shape-dependence in sheared granular packings composed of 
particles of different shapes.   
We find that the shear strength is an increasing function of 
$\eta$ with nearly the same trend for all shapes, 
the differences appearing thus to be of second order compared to $\eta$.  
We also observe a nontrivial behavior of packing fraction which, 
for all our simulated shapes, increases 
with $\eta$ from the random close packing fraction   
for disks, reaches a peak considerably 
higher than that for disks, and subsequently declines as 
$\eta$ is further increased. These findings suggest that 
a low-order description of particle shape accounts for 
the principal trends of packing fraction 
and shear strength. Hence, the effect of second-order shape parameters 
may be investigated by considering different shapes at the same level of $\eta$.}

\begin{document}

\maketitle


The hard-sphere packing is at the heart of various models for the rheology and 
(thermo)dynamical properties of amorphous states of matter including liquids,  
glasses and granular materials \cite{Binder2005,Man2005}. 
Such models reflect  both 
the purely geometrical properties of sphere 
packings, e.g. the order-disorder transition with finite volume 
change \cite{Torquato2000}, and emergent properties arising 
from collective particle interactions, e.g. force chains and arching 
in static piles \cite{Radjai1998}. 
As to non-spherical particle packings, 
rather recent results suggest that such packings exhibit higher shear strength than 
sphere packings \cite{Ouadfel2001,Mirghasemi2002,Nouguier-Lehon2003,
Azema2007,Azema2009,Azema2010,Estrada2011,Azema2012,Nouguier-Lehon2010,Saint-Cyr2011,Szarf2011}, 
and may approach unusually high packing 
fractions \cite{Donev2004,Donev2004a,Man2005,Giao2010}. 
However, a systematic and quantitative investigation of shape-dependence 
is still largely elusive since particle shape characteristics such as elongation, angularity, 
slenderness and nonconvexity are described by distinct groups of  
parameters, and the effect of each parameter is not  
easy to isolate experimentally.

\begin{figure}[tb]
\centering
\includegraphics[width=4cm]{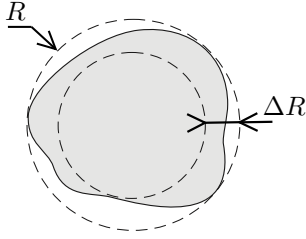}
\caption{An arbitrary particle shape represented by a concentric pair of 
  circumscribing and inscribed circles.}
\label{fig:Ring}
\end{figure}

In order to evaluate the shape-dependence of general  
granular properties such as packing fraction, shear strength and internal structure 
for particles of different shapes, we designed a numerical benchmark test 
that was simulated and analyzed by the members of a collaborative group (CEGEO). 
The idea of this test is that various non-spherical or non-circular 
shapes can be characterized by their degree of 
distortion from a perfectly spherical or circular shape. 
Let us consider an arbitrary 2D shape as sketched in Fig. \ref{fig:Ring}. 
The border of the particle is fully enclosed between two concentric circles: 
a circumscribing circle of radius $R$ and an inscribed circle 
of radius $R-\Delta R$. We define the {\em $\eta$-set} as the set of all shapes with borders 
enclosed between a pair of concentric circles (spheres in 3D), touching both circles    
and having the same ratio
\begin{equation} 
\eta = \frac{\Delta R}{R}. 
\label{eqn:eta}
\end{equation}

\begin{figure}[tbh]
\centering
\includegraphics[width= 8cm]{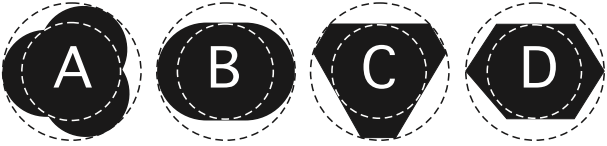}
\caption{Four different shapes belonging to the 
  same $\eta$-set with $\eta=0.4$: trimer (A), 
  rounded-cap rectangle (B), truncated triangle (C), and elongated 
  hexagon (D).}
\label{fig:Shapes}

\end{figure}

Four different particle shapes belonging to the same 
$\eta$-set are shown in Fig. \ref{fig:Shapes}. A non-zero value 
of $\eta$ corresponds to  
non-convexity for A-shape, elongation for B-shape,   
angularity for C-shape, 
and a combination of angularity and elongation for D-shape. 

\begin{figure}[tbh]
\centering
\includegraphics[width= 7cm]{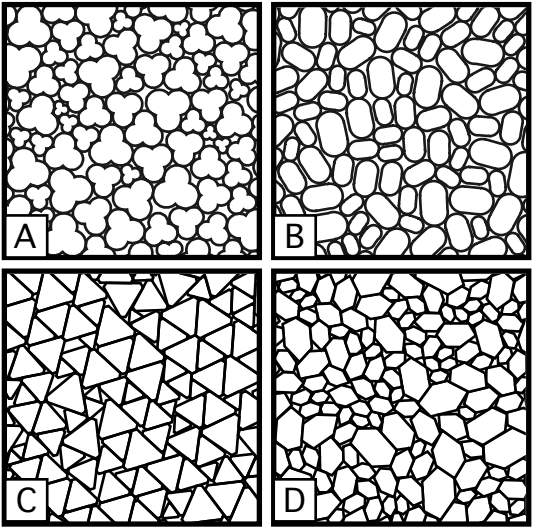}
\caption{Snapshots of the simulated packings in the densest isotropic state for 
  $\eta=0.4$.}
\label{fig:Snaps}
\end{figure}

The parameter $\eta$ is obviously a rough low-order shape 
parameter; see also \cite{Poschel1993a}.   
But, encompassing 
most specific shape parameters, it provides   
a general framework in which shape-dependence 
may be analyzed among particles of very different shapes. 
Within an $\eta$-set, each specific 
shape may further be characterized by 
higher-order parameters. 
The issue that we address in this Letter is   
to what extent the packing fraction and shear strength are controlled by  
$\eta$ and in which respects the behavior depends on 
higher-order shape parameters .    


The benchmark test is based on 
the four shapes of Fig. \ref{fig:Shapes}. 
The A-shape (trimer) is 
composed of  three overlapping disks touching 
the circumscribing circle and with their intersection points lying on 
the inscribed circle; the B-shape (rounded-cap rectangle) is a 
rectangle touching the inscribed circle and juxtaposed with two half-disks 
touching the circumscribing circle; the C-shape (truncated triangle) is a  
hexagon with three sides constrained to touch the inscribed circle and 
all corners on the circumscribing circle; and 
the D-shape (elongated hexagon) is an irregular hexagon with 
two sides constrained to touch the inscribed circle and two corners 
lie on the circumscribing circle. 
The range of geometrically defined values of $\eta$ for a given 
shape (defined by a construction method) has in 
general a lower bound $\eta_0$. For A and  
B, the particle shape changes continuously from 
a disk, so that $\eta_0 = 0$ whereas 
we have $\eta_0=1-\sqrt{3}/2\simeq 0.13$ for C and D.

\begin{figure}[tb]
\centering
\includegraphics[width= 8cm]{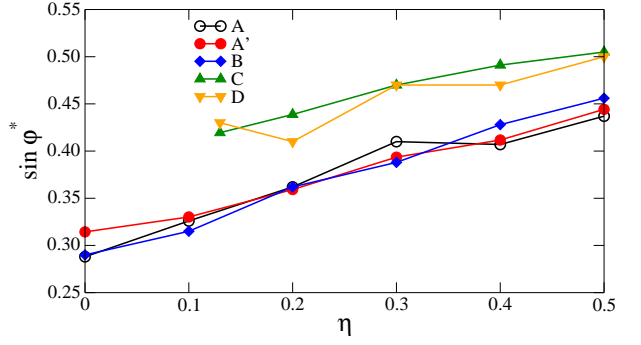}
\caption{Shear strength $\sin \varphi^*$ of packings composed of various 
  particle shapes (see Fig. \ref{fig:Shapes}) as a function of $\eta$.}
\label{fig:Phi}
\end{figure}
\begin{figure}[tb]
\centering
\includegraphics[width= 8cm]{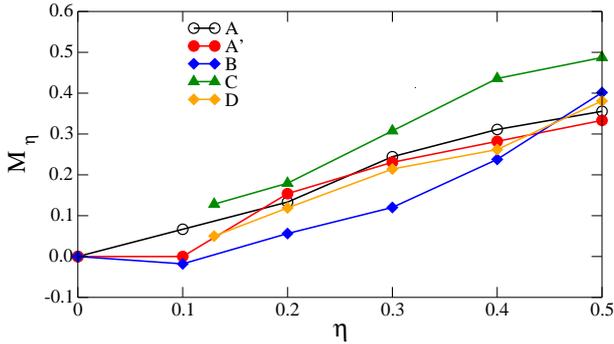}
\caption{Friction mobilization in the steady state as a function of $\eta$ for 
  different particle shapes.}
\label{fig:Friction}
\end{figure}

\begin{figure}[tb]
\centering
\includegraphics[width= 8cm]{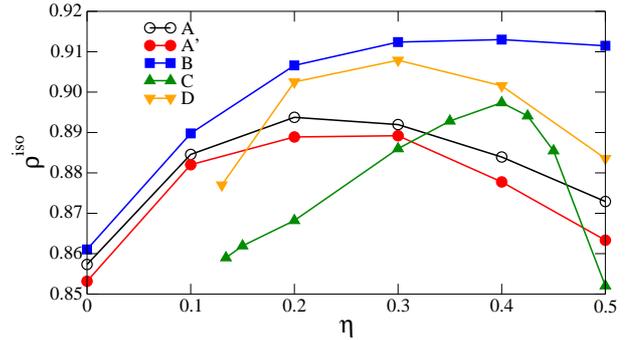}
\caption{Packing fraction in the isotropic state as a function of $\eta$ for 
  different particle shapes.}
\label{fig:Rho_iso}
\end{figure}

Two different discrete element methods (DEM) were used for the simulations: 
contact dynamics (CD) and molecular dynamics (MD).  
In the CD method, the particles are treated as 
perfectly rigid \cite{Radjai2009a} whereas a linear spring-dashpot model was 
used in MD simulations with stiff particles ($k_n/p_0 > 10^3$, where $k_n$ is 
the normal stiffness and $p_0$ refers to the confining pressure) 
\cite{Combe2003}. The trimers  
were simulated by both methods for all values of $\eta$. We refer below as 
A (for CD) and A' (for MD) to these simulations. The packing C was 
simulated by MD whereas the packings B and D were 
simulated by CD. {In CD simulations, the coefficient of 
restitution was set to zero. In MD simulations, the damping 
parameter was taken very close to the critical damping coefficient so that 
the restitution coefficient was also negligibly small \cite{Brilliantov1996b}. 
Note that in quasi-static flow, the relaxation time 
of the particles is short enough (compared to the inverse shear rate) to allow 
for efficient dissipation of kinetic energy in each time step. For this reason, 
in contrast to granular gases, the exact values of the 
damping parameters or restitution coefficients 
have practically no influence on the numerical data analyzed 
below \cite{Chevoir2011}.}  

For each shape, several packings of $5000$ 
particles were prepared with $\eta$ varying  
from 0 to $0.5$. To avoid long-range ordering, 
a size polydispersity was introduced by taking $R$  in 
the range $[R_{min} , R_{max} ]$ with $R_{max} = 3R_{min}$ 
and a uniform distribution of particle volumes.  
A dense packing composed of disks ($\eta=0$) was  
first constructed by means of random 
deposition in a box \cite{Voivret2007}. 
For other values of $\eta$, the same packing was used 
with each disk serving as the circumscribing circle.    
The particle was inscribed with the desired value of $\eta$ and random orientation 
inside the disk. This geometrical step was followed by isotropic 
compaction of the packings inside a rectangular frame.
 The gravity $g$ and friction coefficients between particles 
and with the walls were set to $0$ during compaction in order to
avoid force gradients. Fig. \ref{fig:Snaps}  displays snapshots of the packings for 
$\eta=0.4$ at the end of isotropic compaction\footnote{Animation videos of the 
simulations can be found at www.cgp-gateway.org/ref012.}.

\begin{figure}[tbh]
\centering
\includegraphics[width= 3cm]{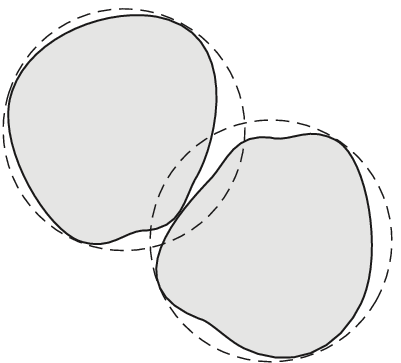}(a)
\includegraphics[width= 3cm]{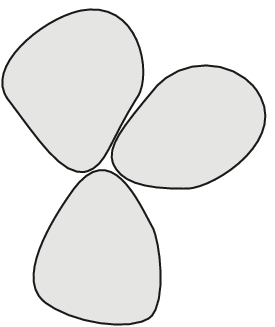}(b)
\caption{Pore volume reduction by (a) overlap between 
  self-porosities; (b) steric pores.}
\label{fig:HillValley}
\end{figure}

The isotropic samples were sheared by applying a slow downward 
velocity on the top wall with 
a constant confining stress acting on the lateral walls.
During shear, the friction coefficient $\mu$  between particles was 
set to $0.5$ and to $0$ with the walls. 
The shear strength is characterized by the 
internal angle of friction $\varphi$ defined by 
\begin{equation}
\sin \varphi = \frac{\sigma_1 - \sigma_2}{\sigma_1 + \sigma_2},
\end{equation}
where  the subscripts $1$ and $2$ refer to the principal stresses. 
$\sin \varphi$ increases rapidly from zero   
to a peak value before relaxing to a constant material-dependent 
value $\sin \varphi^*$, which defines the shear strength at large  
strain at a steady stress state.


Figure \ref{fig:Phi} shows the  dependence of $\sin \varphi^*$ 
with respect to $\eta$ for our different shapes.     
Remarkably, $\sin \varphi^*$ increases with  
$\eta$ at the same rate  for all shapes. 
The data nearly coincide between the  A and B shapes, on the one hand,  
and between C and D shapes, on the other hand. This suggests  that 
nonconvex trimers and  
rounded-cap rectangles, in spite of their very different shapes, belong  to 
the same family (rounded shapes). In the same way, the truncated triangles and 
elongated hexagons seem to belong to the family of angular particles and exhibit 
a shear strength slightly above that of rounded shapes.  
Note also that the results are robust with respect to the numerical approach as 
the packings A and A' were simulated by two different methods.

\begin{figure}[tbh]
\centering
\includegraphics[width= 8cm]{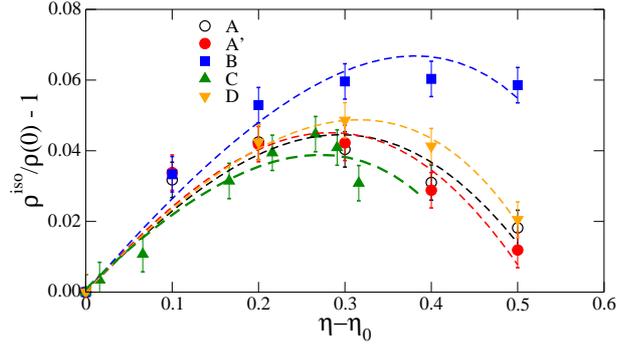}
\caption{Normalized packing fractions  fitted by 
  Eq. (\ref{eqn:rho}).}
\label{fig:Rho_iso_fits}
\end{figure}

The increase of shear strength with $\eta$ may be attributed to  
the increasing frustration of particle rotations as the shape 
deviates from { a} disk \cite{Estrada2008,Estrada2011}. Since the particles may interact  
at two or three contact points (A-shape) or through side-to-side 
contacts (shapes B, C and D), the kinematic constraints 
increase with $\eta$ and frustrate the particle displacements by rolling. 
The restriction of rolling leads to enhanced  
role of friction in the mechanical equilibrium and relative sliding of particles 
during deformation. A related static quantity is 
the mean friction mobilization defined by  
$M= \langle f_t /(\mu f_n) \rangle$, where $f_t$ is the magnitude of the friction 
force, $f_n$ is the normal force, and the average is taken over  
all force-bearing contacts in the system. 

To evaluate the effect of particle shape,  we consider the parameter   
\begin{equation}
M_\eta = \frac{M(\eta)}{M(\eta=0)}-1
\end{equation}
as a function of $\eta$ for different shapes, where 
$M(\eta=0)$ is the friction mobilization for 
circular particles. Fig. \ref{fig:Friction} shows that $M_\eta$ is a globally  
increasing function of $\eta$ for all shapes. 
The parameter $\eta$ appears also in this respect to account for 
the global trend of friction mobilization, and the differences observed in Fig. \ref{fig:Friction}  
among different shapes are rather of second order.      

{We also observe that the proportions of double and triple contacts for 
A-shape packings and the proportion of side-to-side contacts for 
other shapes increase with $\eta$.  
For noncircular particles, one should distinguish 
the coordination number $Z$, defined 
as the mean number of contacting neighbors per particle, from 
the ``contact coordination number'' $Z_c$ defined as the mean number of 
contacts per particle. Obviously, for the calculation of both $Z$ and $Z_c$ 
only the force-bearing contacts and 
non-floating particles are taken into account \cite{Combe2011}.    
We have $Z=Z_c\simeq 4$ for the 
disks in the initial state  prepared with $\mu=0$. 
This value corresponds to an 
isostatic state in which one expects $Z=2N_f$, where $N_f$  
is the number of degrees of freedom of a particle \cite{Agnolin2007}. 
For frictionless disks, we have $N_f=2$ (two translational degrees of freedom), 
leading to $Z=4$. 
For noncircular shapes, we have $N_f=3$ since the rotational degrees 
of freedom take part in the mechanical equilibrium of the particles.   
Hence, if isostaticity holds also for noncircular frictionless  particles, 
we expect $Z=6$. We observe instead $Z<5$ for all our packings. 
However, we find $Z_c \simeq 6$ for $\eta \ne 0$ if   
each side-to-side contact is counted twice, representing two independent 
constraints. This result is consistent with the isostatic nature of 
a packing of frictionless noncircular particles and shows that 
the packings of noncircular shape are not under-constrained as 
previously suggested \cite{Donev2007}.  For $\mu = 0.5$, the 
packings are no more isostatic and   
$Z$ and $Z_c$ vary only slightly with $\eta$ with values 
in the range $3$ to $4$ for $Z$ and in the range 
$4$ to $5$ for $Z_c$ in the course of shearing.}


We now focus on the packing fraction which crucially depends on 
particle shape.  Fig. \ref{fig:Rho_iso} shows the packing fraction $\rho^{iso}$ 
in the initial isotropic state   
as a function of $\eta$. We observe a nontrivial  
behavior for all particle shapes: the packing fraction increases 
with $\eta$,  
passes by a peak depending on each specific shape 
and subsequently declines. For the B-shape a sharp decrease 
of $\rho^{iso}$ occurs beyond $\eta=0.5$ as was shown in \cite{Azema2010}. 

This unmonotonic behavior of packing fraction was   
observed by experiments and numerical simulations for 
spheroids as a function of their aspect ratio    
\cite{Williams2003,Donev2004,Donev2004a,Man2005,Donev2007,Sacanna2007}. 
The decrease of the packing fraction  
is attributed to the excluded-volume effect that  
prevails at large aspect ratios and leads to increasingly larger pores 
which cannot be filled by the particles \cite{Williams2003}. 
The observation of this unmonotonic behavior as a function of $\eta$ 
for different shapes indicates that it is a generic property depending only on 
deviation from circular shape. This behavior may thus be explained 
from general considerations involving the parameter $\eta$ but with variations 
depending on second-order shape characteristics.

A plausible second-order parameter is   
\begin{equation}
\nu=\frac{V_p}{\pi R^2}, 
\end{equation}
where $V_p$ is the 
particle volume in 2D. Its complement $1-\nu$ is the ``self-porosity'' 
of a particle, i.e. the unfilled volume fraction inside the circumscribing circle.  
Keeping the radius $R$ of the circumscribing circle 
constant, $\rho^{iso}= V_p/V$ varies with $\eta$ as a result of 
the relative changes of  $V_p$ and the mean volume 
$V$ per particle. The free (pore) volume per particle is 
$V_f=V - V_p$. 

At $\eta=0$, the free volume $V_f$ is only composed of 
{\em steric voids}, i.e. voids between  three or more particles,  
and the packing fraction is given by $\rho(0) = \pi R^2/V(0)$. 
For $\eta > 0$, the void patterns are more complex    
but can be described by considering the generic shape of particles 
belonging to a given $\eta$-set. The borders of a particle 
involve ``hills'', which are the parts touching the circumscribing circle, 
and ``valleys'' touching the inscribed circle. 
The volume $V$ per particle 
varies with $\eta$ by two mechanisms. 
First, the hills of a particle may partially fill the valleys of 
a neighboring particle; Fig. \ref{fig:HillValley}(a). 
Secondly, the steric voids between the hills shrink as $\eta$ increases 
due to the increasing local curvature of the touching 
particles; Fig. \ref{fig:HillValley}(b). To represent this excess or loss of pore 
volume due to the specific jamming configurations induced by 
particle shapes, we introduce the function $h(\eta)$ by 
setting  
\begin{equation}
V(\eta) =V(\eta_0) - \pi R^2 h(\eta), 
\end{equation}
with $h(\eta_0)=0$. With these assumptions, the packing fraction is expressed as   
\begin{equation}
\rho(\eta) = \frac{\nu(\eta)\rho(\eta_0)}{1 - h(\eta)\rho(\eta_0) }.
\label{eqn:rho}
\end{equation}  

The function $\nu(\eta)$ is known for each shape 
but $h(\eta)$ needs to be estimated. 
A second-order polynomial approximation 
\begin{equation}
h(\eta)= \alpha (\eta - \eta_0) + \beta (\eta-\eta_0)^2
\end{equation}
together with Eq. \ref{eqn:rho} allows us to recover the correct trend and to fit the 
data as shown in Fig. \ref{fig:Rho_iso_fits}. The 
error bars represent the variability at $\eta_0$ assumed to 
be the same for all other values of $\eta$. 
The parameter $\alpha$ ensures the increase of packing fraction 
with $\eta$ at low values of the latter and it basically reflects   
the shrinkage of steric pores (Fig. \ref{fig:HillValley}(b)) whereas   
$\beta$ accounts for the overlap between circumscribing 
circles (Fig. \ref{fig:HillValley}(a))) and is responsible for 
the subsequent decrease of the packing fraction.

The fitting parameters in Fig. \ref{fig:Rho_iso_fits} 
are $\alpha \simeq$ 1.30, 1.29, 1.14, 1.17 
  and $\beta \simeq$ 1.23, 1.20, 0.23, 0.20 
for C, A, D and B shapes, respectively with increasing peak value. 
Note that the values of $\beta$ are considerably 
smaller for B and D that have an elongated aspect and for which the 
overlapping of self-porosities prevails as compared to 
A and C for which the shrinkage of the initial pores is more important.  


In summary, our benchmark simulations show that 
a low-order shape parameter $\eta$, describing deviation 
with respect to circular shape, controls 
to a large extent both the shear strength and 
packing fraction  
of granular media composed of noncircular particles in 2D. 
The shear strength is roughly linear 
in $\eta$ whereas the packing fraction is unmonotonic. 
Our simple model for this  unmonotonic behavior is 
consistent with the numerical data for all shapes. 
It is governed by a first-order term in $\eta$ for the shrinkage 
of the initial steric pores and a second-order term in $\eta$ for the 
creation of large pores by shape-induced steric pores.    
The effect of higher-order shape parameters may be analyzed also in 
this framework in terms of differences in packing fraction 
and shear strength among various shapes belonging to the 
same $\eta$-set. An interesting issue to be addressed in future is whether 
a generic second-order parameter accounting for such differences exists. 
{Another aspect that merits further investigation is the joint effects of 
size polydispersity and particle shape. The shear strength is 
independent of particle size polydispersity as a result of the capture of 
force chains by the class of larger particles \cite{Voivret2009a}. But 
the packing fraction and force and contact anisotropy depend 
on both shape and polydispersity.}      

\acknowledgments
We thank B. Cambou and F. Nicot for stimulating discussions. 
We also acknowledge financial support of the French government through the 
program PPF CEGEO.     

\bibliographystyle{eplbib} 

\begin{thebibliography}{10}
\expandafter\ifx\csname url\endcsname\relax\def\url#1{\texttt{#1}}\fi

\bibitem{Binder2005}
\Name{Binder K. \and Kob W.} \Book{Glassy materials and disordered solids}
  (World Scientific) 2005.

\bibitem{Man2005}
\Name{Man W., Donev A., Stillinger F., Sullivan M., Russel W., Heeger D.,
  S.Inati, Torquato S. \and Chaikin P.} \REVIEW{Phys. Rev.
  Lett.}{94}{2005}{198001}.

\bibitem{Torquato2000}
\Name{Torquato S., Truskett T.~M. \and Debenedetti P.~G.} \REVIEW{Phys. Rev.
  Lett.}{84}{2000}{2064}.

\bibitem{Radjai1998}
\Name{Radjai F., Wolf D.~E., Jean M. \and Moreau J.} \REVIEW{Phys. Rev.
  Letter}{80}{1998}{61}.

\bibitem{Ouadfel2001}
\Name{Ouadfel H. \and Rothenburg L.} \REVIEW{Mechanics of
  Materials}{33}{2001}{201}.

\bibitem{Mirghasemi2002}
\Name{Mirghasemi A., Rothenburg L. \and Maryas E.}
  \REVIEW{Geotechnique}{52}{2002}{N 3, 209}.

\bibitem{Nouguier-Lehon2003}
\Name{Nouguier-Lehon C., Cambou B. \and Vincens E.} \REVIEW{Int. J. Numer.
  Anal. Meth. Geomech.}{27}{2003}{1207}.

\bibitem{Azema2007}
\Name{Az\'ema E., Radjai F., Peyroux R. \and Saussine G.} \REVIEW{Phys. Rev.
  E}{76}{2007}{011301}.

\bibitem{Azema2009}
\Name{Az\'ema E., Radjai F. \and Saussine G.} \REVIEW{Mechanics of
  Materials}{41}{2009}{721}.

\bibitem{Azema2010}
\Name{Az\'ema E. \and Radjai F.} \REVIEW{Phys. Rev. E}{81}{2010}{051304}.

\bibitem{Estrada2011}
\Name{Estrada N., Az\'ema E., Radjai F. \and Taboada A.} \REVIEW{Phys. Rev.
  E}{84}{2011}{011306}.

\bibitem{Azema2012}
\Name{Az\'ema E. \and Radjai F.} \REVIEW{Phys. Rev. E}{85}{2012}{031303}.

\bibitem{Nouguier-Lehon2010}
\Name{Nouguier-Lehon C.} \REVIEW{Comptes Rendus M\'ecanique}{338}{2010}{587}.

\bibitem{Saint-Cyr2011}
\Name{Saint-Cyr B., Delenne J.-Y., Voivret C., Radjai F. \and Sornay P.}
  \REVIEW{Phys. Rev. E}{84}{2011}{041302}.

\bibitem{Szarf2011}
\Name{Szarf K., Combe G. \and Villard P.} \REVIEW{Powder
  Technology}{208}{2011}{279}.

\bibitem{Donev2004}
\Name{Donev A., Stillinger F., Chaikin P. \and Torquato S.} \REVIEW{Phys. Rev.
  Lett.}{92}{2004}{255506}.

\bibitem{Donev2004a}
\Name{Donev A., Cisse I., Sachs D., Variano E., Stillinger F., Connelly R.,
  Torquato S. \and Chaikin P.} \REVIEW{Science}{303}{2004}{990}.

\bibitem{Giao2010}
\Name{Jiao Y., Stillinger F.~H. \and Torquato S.} \REVIEW{Phys. Rev.
  E}{041304}{2010}{1}.

\bibitem{Poschel1993a}
\Name{P\"oschel T. \and Buchholtz V.} \REVIEW{Phys. Rev.
  Lett.}{71}{1993}{3963}.

\bibitem{Radjai2009a}
\Name{Radjai F. \and Richefeu V.} \REVIEW{Mechanics of
  Materials}{41}{2009}{715}.

\bibitem{Combe2003}
\Name{Combe G. \and Roux J.-N.} \Book{Discrete numerical simulation,
  quasistatic deformation and the origins of strain in granular materials} in
  proc. of \Book{Deformation Characteristics of Geomaterials}, edited by
  \Name{et~al. D.~B.} 2003 pp. 1071--1078.

\bibitem{Brilliantov1996b}
\Name{Brilliantov N.~V., Spahn F., Hertzsch J.-M. \and P\"oschel T.}
  \REVIEW{Phys. Rev. E}{53}{1996}{5382}.

\bibitem{Chevoir2011}
\Name{Roux J.-N. \and Chevoir F.} \Book{Dimensional analysis and control parameter} 
in book. of \Book{Discrete-element Modeling of Granular Materials (ISTE-Wiley)}, edited by
  \Name{F.Radjai and F.Dubois.} 2011 Ch.~8 pp. 223--253.

\bibitem{Voivret2007}
\Name{Voivret C., Radjai F., Delenne J.-Y. \and Youssoufi M. S.~E.}
  \REVIEW{Phys. Rev. E}{76}{2007}{021301}.

\bibitem{Estrada2008}
\Name{Estrada N., Taboada A. \and Radjai F.} \REVIEW{Phys. Rev.
  E}{78}{2008}{021301}.

\bibitem{Combe2011}
\Name{Combe G. \and Roux J.-N.} \Book{Construction of granular assemblies under
  static loading} in book. of \Book{,Discrete-element Modeling of Granular Materials (ISTE-Wiley)}, edited by
  \Name{F.Radjai and F.Dubois.} 2011 Ch.~6 pp. 153--180.

\bibitem{Agnolin2007}
\Name{Agnolin I. \and Roux J.-N.} \REVIEW{Phys; Rev. E}{76}{2007}{061302}.

\bibitem{Donev2007}
\Name{Donev A., Connelly R., Stillinger F. \and Torquato S.} \REVIEW{Phys. Rev.
  E}{75}{2007}{051304}.

\bibitem{Williams2003}
\Name{Williams S. \and Philipse A.} \REVIEW{Phys. Rev. E}{67}{2003}{051301}.

\bibitem{Sacanna2007}
\Name{Sacanna S., Rossi L., Wouterse A. \and Philipse A.} \REVIEW{Journal of
  Physics}{19}{2007}{376108}.

\bibitem{Voivret2009a}
\Name{Voivret C., Radjai F., Delenne J.-Y. \and El~Youssoufi M.~S.}
  \REVIEW{Phys. Rev. Lett.}{102}{2009}{178001}.

\end{thebibliography}

\end{document}